# Measurement-Based Assessment of Decadal Inertia and Primary Frequency Response Trends in the US Eastern Interconnection

**S. DULAL[1], M. BALDWIN[2], N. GHIASI[2], M. OLAMA[3], A.R. EKTI[3], N. M. STENVIG[3], Y. LIU[1,3]**
**[1]The University of Tennessee, Knoxville**
**[2]Dominion Energy**
**[3]Oak Ridge National Laboratory**
**USA**

## SUMMARY

Frequency performance in a large power interconnection depends on more than one dynamic response. Power system inertia affects the initial rate-of-change-of-frequency (RoCoF) immediately after a disturbance. Primary frequency response (PFR) helps arrest and stabilize the frequency excursion after the disturbance begins to develop. In this paper, PFR is evaluated using the beta metric, which represents the aggregate frequency response observed from disturbance events. Since inertia and PFR describe different parts of the frequency response, they should be studied together when evaluating long-term changes in system performance.

This paper presents a measurement-based analysis of long-term inertia and PFR/beta trends in the U.S. Eastern Interconnection (EI). The analysis uses synchronized wide-area frequency measurements from the Frequency Monitoring Network (FNET/GridEye) and confirmed generation-trip disturbance records. A representative EI frequency is obtained from geographically dispersed frequency disturbance recorders, and screened disturbance events are used for inertia and PFR/beta trend estimation.

Effective inertia is estimated from the early post-disturbance RoCoF response. A 3-second RoCoF window is used for the inertia calculation. Quarterly inertia estimates are used so that seasonal variation can be observed while preserving a reasonable number of events in each estimation period. The PFR/beta metric is estimated annually using a least-squares relationship between disturbance size and frequency deviation. Since beta is sensitive to event conditions, filtering is applied using starting frequency and disturbance size to improve comparability between years.

sdulal@vols.utk.edu
This manuscript has been authored by UT-Battelle, LLC, under contract DE-AC05-00OR22725 with the US Department of Energy (DOE). The US government retains and the publisher, by accepting the article for publication, acknowledges that the US government retains a nonexclusive, paid-up, irrevocable, worldwide license to publish or reproduce the published form of this manuscript, or allow others to do so, for US government purposes. DOE will provide public access to these results of federally sponsored research in accordance with the DOE Public Access Plan (https://www.energy.gov/doe-public-access-plan).

The results show that EI inertia is strongly seasonal and non-monotonic. The trend does not show a uniform long-term decline. Instead, it includes periods of reduction, relatively steady behavior, and modest recovery. This suggests that effective EI inertia is shaped by several offsetting factors and is not explained by renewable integration alone.

Generation-mix data provide useful context for this result. Over the study period, coal generation decreased, while natural gas, wind, and solar generation increased. A simple expectation might be that higher inverter-based resource penetration would cause a strong decline in effective inertia. However, the measured EI inertia trend does not support such a simple interpretation. Increased gas-fired generation, continued operation of synchronous resources, seasonal commitment patterns, and operating conditions appear to have partially offset the expected inertia reduction.

The filtered PFR/beta trend gives a different view of EI frequency performance. After event conditioning, the annual beta trend generally increases from the mid-2010s to the early 2020s, with year-to-year variation in later years. This suggests that the exhibited aggregate PFR has changed over time and may be related to changing resource participation and governor-responsive behavior.

The joint interpretation shows that EI frequency performance has not changed in a simple one-directional way. Effective inertia did not decline uniformly, and the filtered PFR/beta trend does not directly follow the inertia trend. This means that early inertial response and broader aggregate PFR have evolved differently. Overall, the study shows that inertia and PFR/beta provide a more complete understanding of long-term EI frequency performance when interpreted together. Future work will expand the qualified event database and test the sensitivity of beta trends to filtering criteria.

sdulal@vols.utk.edu

## 1. INTRODUCTION

Frequency performance in a large power interconnection is shaped by multiple dynamic attributes that act over different stages of a disturbance. Among them, inertia and primary frequency response (PFR) are two closely related but distinct mechanisms. Inertia governs the initial rate-of-change-of-frequency (RoCoF) immediately after an active power imbalance, while PFR helps arrest and stabilize the resulting frequency excursion through generator governor action [1]. In traditional synchronous-generator-dominated systems, both mechanisms were largely provided by conventional generation fleets. However, as power systems transition toward more diverse resource mixes, long-term changes in inertia and PFR have become increasingly important for understanding frequency behavior at the interconnection level.

The U.S. Eastern Interconnection (EI) has experienced substantial changes in generation composition during the past decade. Coal-fired generation has declined, while natural gas, wind, and solar generation have increased [2]. These changes complicate the interpretation of interconnection-wide frequency behavior. On one hand, increasing inverter-based resources (IBRs) integration raises concerns about declining effective inertia because IBRs do not inherently contribute synchronous rotational energy. On the other hand, replacement of coal units by gas-fired units, continued operation of synchronous resources, changing unit-commitment patterns, and evolving control capabilities may alter the observed inertial and primary response characteristics in different ways. As a result, a single metric is often insufficient to explain how EI frequency performance has evolved over time.

Measurement-based evaluation provides a practical way to assess these changes because the observed disturbance response inherently reflects the composite behavior of the operating system. However, extracting consistent interconnection-level metrics from field data is not straightforward. In large-scale systems such as the EI, local oscillations, propagation delays, non-uniform sensor distribution, and regional frequency differences can distort the use of individual measurements. Prior EI studies [3, 4] using Frequency Monitoring Network (FNET/GridEye) data have shown that a representative median frequency is more suitable for interconnection-level analysis than relying on localized traces. They have also shown that inertia estimation depends strongly on proper event-start detection and appropriate RoCoF window selection, while PFR estimation through the beta metric is sensitive to event conditions such as pre-disturbance frequency and disturbance MW size. These practical considerations are essential for obtaining fair and meaningful trend comparisons across years.

Although inertia trend analysis in [3] and PFR studies [4] have each been investigated using synchronized field measurements, they are usually treated separately. This leaves an important gap in long-term interpretation. Inertia mainly characterizes the system's early electromechanical buffering capability, whereas PFR reflects a broader aggregate response shaped by governor deadbands, operating state, event size, load behavior, and resource mix. Therefore, these two quantities should not be interpreted as interchangeable. Instead, their joint behavior can provide a more complete picture of how EI frequency performance is evolving. If inertia and PFR move in the same direction, they may indicate a broader structural shift in frequency performance. If they diverge, the divergence itself becomes informative, suggesting that changes in system-wide response capability are not explained by effective inertia alone.

Motivated by this need, this paper presents a measurement-based joint trend analysis of inertia and PFR in the U.S. eastern interconnection. Historical disturbance data from the FNET/GridEye synchronized measurement network and confirmed event information are used to construct a decade-long view of EI frequency behavior. The analysis first develops an interconnection-level representative frequency from wide-area measurements, then estimates quarterly inertia using RoCoF-based disturbance analysis and finally evaluates annual PFR trends through a filtered beta framework designed to improve event comparability. Fuel-mix information is used to support interpretation of the inertia trend, while the event-conditioning characteristics of beta are explicitly considered in the PFR trend analysis.

The main contributions of this paper are as follows. First, it develops a unified field-measurement-based framework for jointly evaluating long-term EI inertia and PFR trends from generation trip events. Second, it combines median-frequency preprocessing, event-start detection, RoCoF window selection, and event comparability screening into a consistent methodology suitable for interconnection-level trend analysis. Third, it provides a joint interpretation of inertia and beta trends, showing how these two complementary metrics can reveal different aspects of the evolution of EI frequency performance under a changing generation portfolio.

The remainder of this paper is organized as follows. Section 2 describes the data source and preprocessing procedures. Section 3 presents the unified estimation methodology for inertia and beta. Section 4 gives the trend results and their joint interpretation for the EI. Section 5 concludes the paper.

## 2. FIELD MEASUREMENTS AND PREPROCESSING

### 2.1 FNET/GridEye and Disturbance-Event Data

The analysis in this paper is based on synchronized wide-area frequency measurements collected by the Frequency Monitoring Network (FNET/GridEye), a distribution-level phasor measurement network deployed across North America. Fig. 1 shows the deployment of FNET/GridEye sensors across the United States. These frequency disturbance recorders provide time-synchronized measurements of frequency, voltage magnitude, and phase angle from geographically dispersed locations, making it possible to observe interconnection-level frequency dynamics during large disturbances [5]. In this study, FNET/GridEye data stored at the University of Tennessee, Knoxville and Oak Ridge National Laboratory are used to obtain event frequency measurements, while North American Electric Reliability Corporation (NERC) records are used to obtain the corresponding event size in MW. The event set consists of historical generation-trip disturbances in the EI selected from the study period for long-term trend analysis.

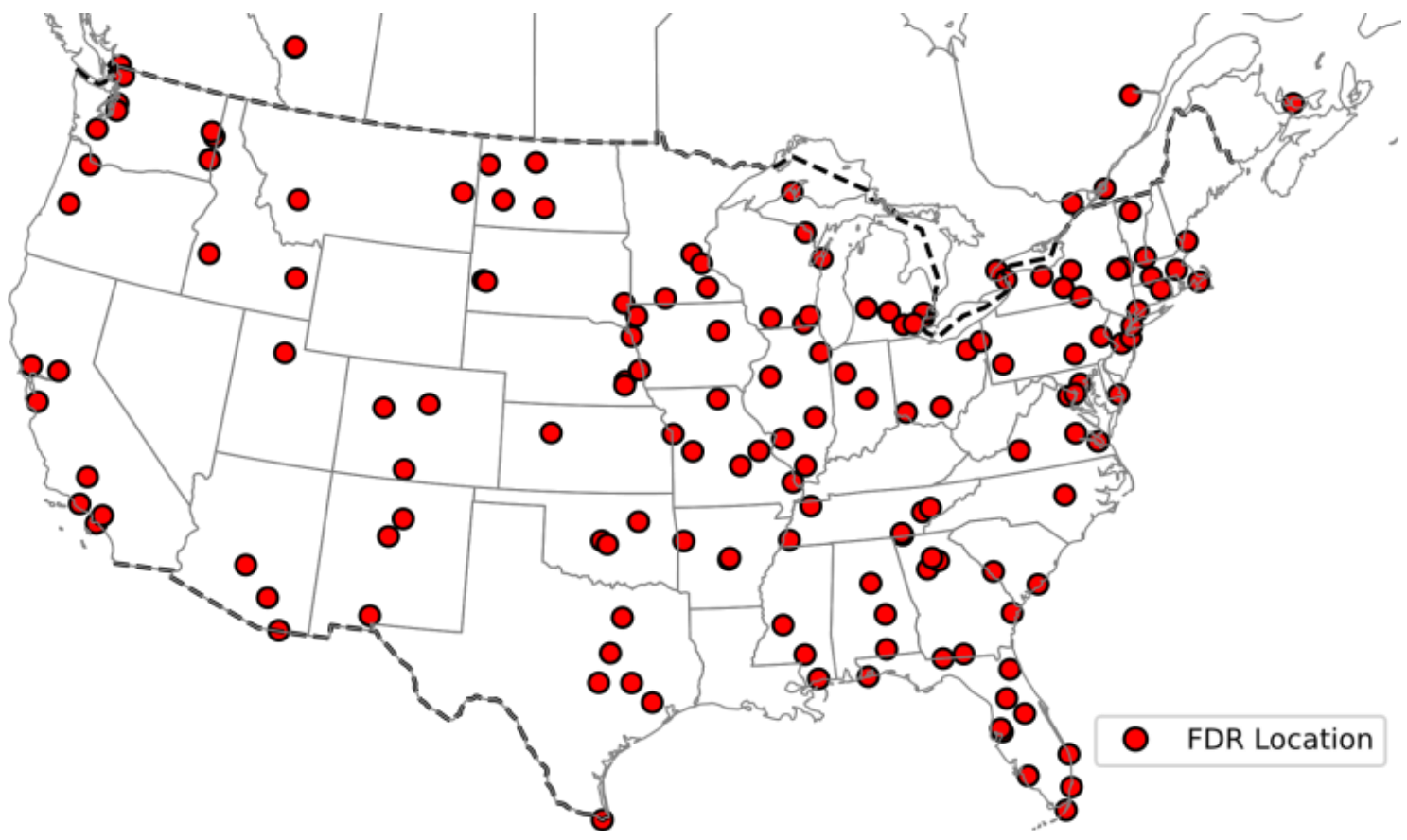

Figure 1 : Deployment of FNET/GridEye sensors in the U.S.

### 2.2 Interconnection-Level Representative Frequency

Because the EI covers a vast geographical area, local frequency traces can differ substantially during the first few seconds after a disturbance. Regional oscillations, propagation delays, and location-specific dynamics may cause measurements near the event location to deviate significantly from measurements in distant areas. Therefore, a single local trace is not sufficient to represent the overall response of the interconnection. Prior measurement-based studies using FNET/GridEye have shown that an interconnection-level representative frequency is needed for reliable trend analysis in large systems such as the EI. This work uses the median frequency across all active sensors at each sampling instant as the representative EI frequency, similar to [3]. This makes it better suited for interconnection-level analysis, where the objective is to capture the central tendency of the EI frequency trajectory rather than over-emphasize regional behavior. The representative median frequency is calculated as follows:

$$f_{median} = \begin{cases} F\left[\frac{m+1}{2}\right], & m \text{ is odd} \\ \frac{F\left[\frac{m}{2}\right] + F\left[\frac{m}{2}+1\right]}{2}, & m \text{ is even} \end{cases} \quad (1)$$

where $m$ is the number of active sensors at a given sampling instant and $F$ is the ordered list of frequency measurements reported by those sensors. The resulting $f_{\text{median}}$ is used throughout the paper for both inertia and PFR estimation.

## 3. UNIFIED ESTIMATION METHODOLOGY

With the representative interconnection frequency, the next step is to estimate the two frequency-performance metrics considered in this paper: inertia and PFR. Fig. 2 presents the overall workflow of the proposed joint estimation framework. Starting from synchronized FNET/GridEye frequency measurements and confirmed disturbance-event information, the method proceeds through two parallel estimation paths: i.) an inertia-estimation path based on event start-point detection, RoCoF calculation, time-window selection, and quarterly least-squares fitting, and ii.) a PFR path based on frequency deviation, event conditioning, and beta estimation. The resulting inertia and beta trends are then jointly interpreted to characterize the long-term evolution of frequency performance in EI.

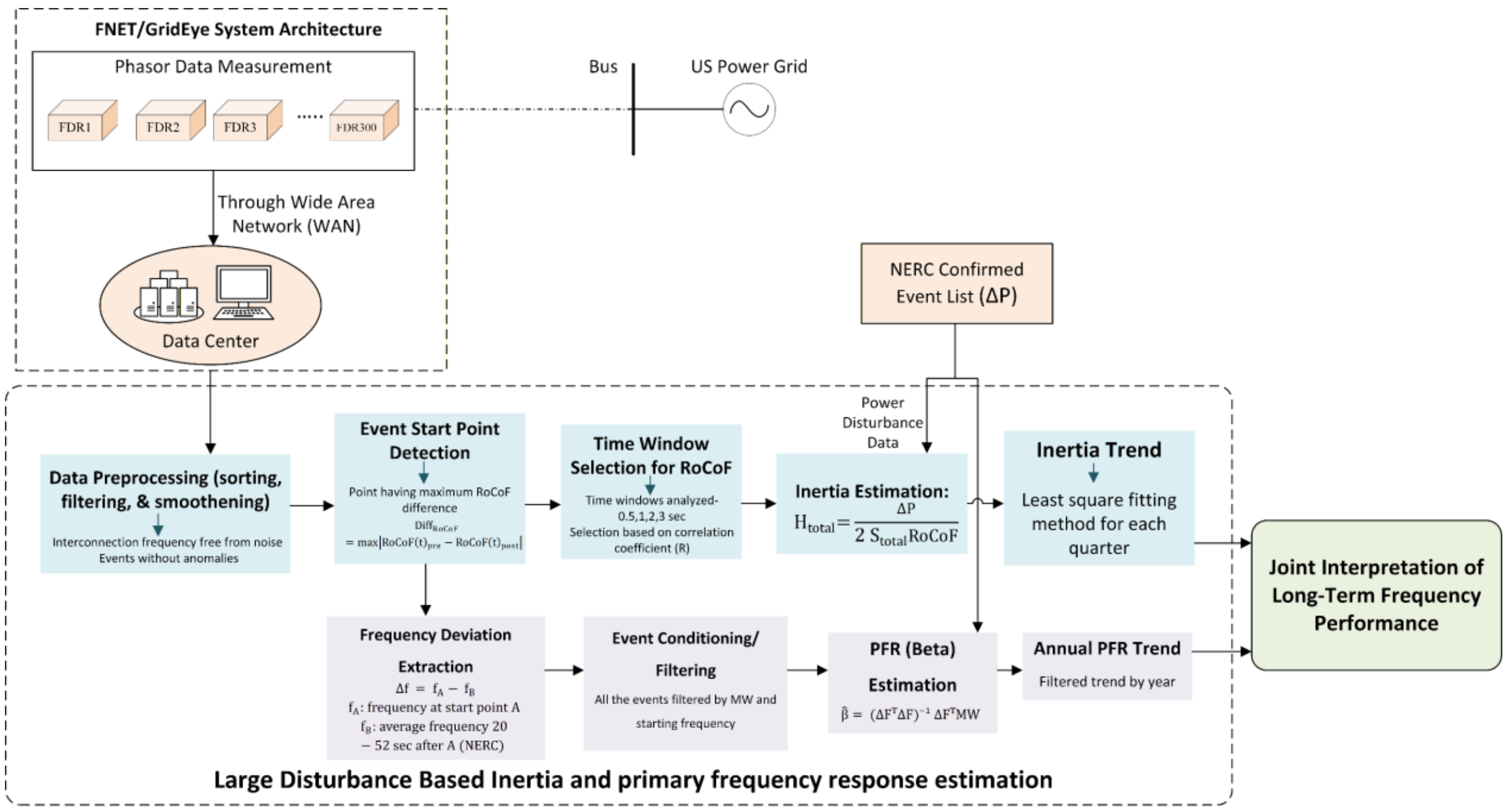

Figure 2: Proposed measurement-based framework for estimating long-term inertia and PFR trends

### 3.1 Event Start-Point Detection

A reliable disturbance start point is essential for both inertia estimation and PFR evaluation. For interconnection-level analysis, the event time recorded by the first sensor that detects the disturbance is not always suitable, because the representative EI frequency used in this study is the median frequency rather than an individual local trace. Similarly, estimating the event start point by averaging the pre-disturbance frequency over a fixed interval may not provide an accurate basis for RoCoF calculation. Therefore, this work identifies the disturbance start point directly from the median-frequency trajectory by locating the most significant change in RoCoF around the event. The event start point is selected using the maximum difference between the pre- and post-disturbance RoCoF values:

$$Diff_{RoCoF} = max\left|RoCoF(t)_{pre} - RoCoF(t)_{post}\right| \quad (2)$$

where $RoCoF(t)_{pre}$ is calculated using the frequency at time $t$ and the frequency at $t - \Delta t$, while $RoCoF(t)_{post}$ is calculated using the frequency at time $t$ and the frequency at $t + \Delta t$. Here, $\Delta t$ denotes the short interval used to compare the local pre- and post-event slopes around each candidate point. The point with the largest value of $Diff_{RoCoF}$ is chosen as the disturbance start point. This approach is intended to capture the turning point of the interconnection-level frequency response more accurately than relying on the first local sensor trigger.

### 3.2 RoCoF Calculation and Time-Window Selection

A key issue in large-interconnection inertia estimation is the choice of the RoCoF calculation window. In a small system or local area, a short time window may be sufficient because the disturbance onset is sharp and the local frequency decline appears almost immediately. In the EI, however, the interconnection-level median frequency often changes more gradually in the first second after the event because inertial support from remote areas takes time to propagate to the disturbance location [6]. This causes the median-frequency response to be initially flatter than nearby regional traces. As a result, an excessively short window may under-represent the interconnection response and be more sensitive to noise, whereas an excessively long window may begin to include participation from governors. This tradeoff is also illustrated conceptually in Fig. 3. The inertia-estimation window covers the early post-disturbance RoCoF interval. After that, governor action begins once the frequency deviation exceeds the governor deadband threshold.

For a selected window length $T$, RoCoF is calculated as:

$$RoCoF_T = \frac{(f_{start} - f_{start+T})}{T} \tag{3}$$

where $f_{start}$ is the median frequency immediately before the detected event start point and $f_{start+T}$ is the median frequency $T$ seconds later. In this paper, a 3-second RoCoF window is adopted for EI inertia estimation, following the correlation-based window-selection analysis established in the prior measurement-based EI inertia work [3]. That analysis showed that the correlation between event MW and RoCoF increases with time-window length and becomes nearly saturated after 3 seconds. This makes 3 seconds a practical compromise between interconnection-level representation and limited participation from governor.

### 3.3 Inertia Estimation from Disturbance Events

The inertia evaluation in this paper is based on the early post-disturbance frequency response, where the frequency deviation remains small and the damping term can be neglected. Under this assumption, the effective interconnection inertia can be related to disturbance size and RoCoF through the simplified swing-equation-based relation [7]:

$$M_{sys} = \frac{\Delta P}{2 \cdot RoCoF_{sys}} \tag{4}$$

where $M_{sys}$ is the effective system inertia in MVA·s, $\Delta P$ is the disturbance size in MW, and $RoCoF_{sys}$ is the interconnection-level rate of change of frequency. Accordingly, the inertia estimate in this paper should be interpreted as an effective disturbance-based interconnection inertia, rather than as a direct inventory of only rotating synchronous kinetic energy. As such, the result reflects the aggregate effect of synchronous machines, loads, and other fast dynamic responses present in EI.

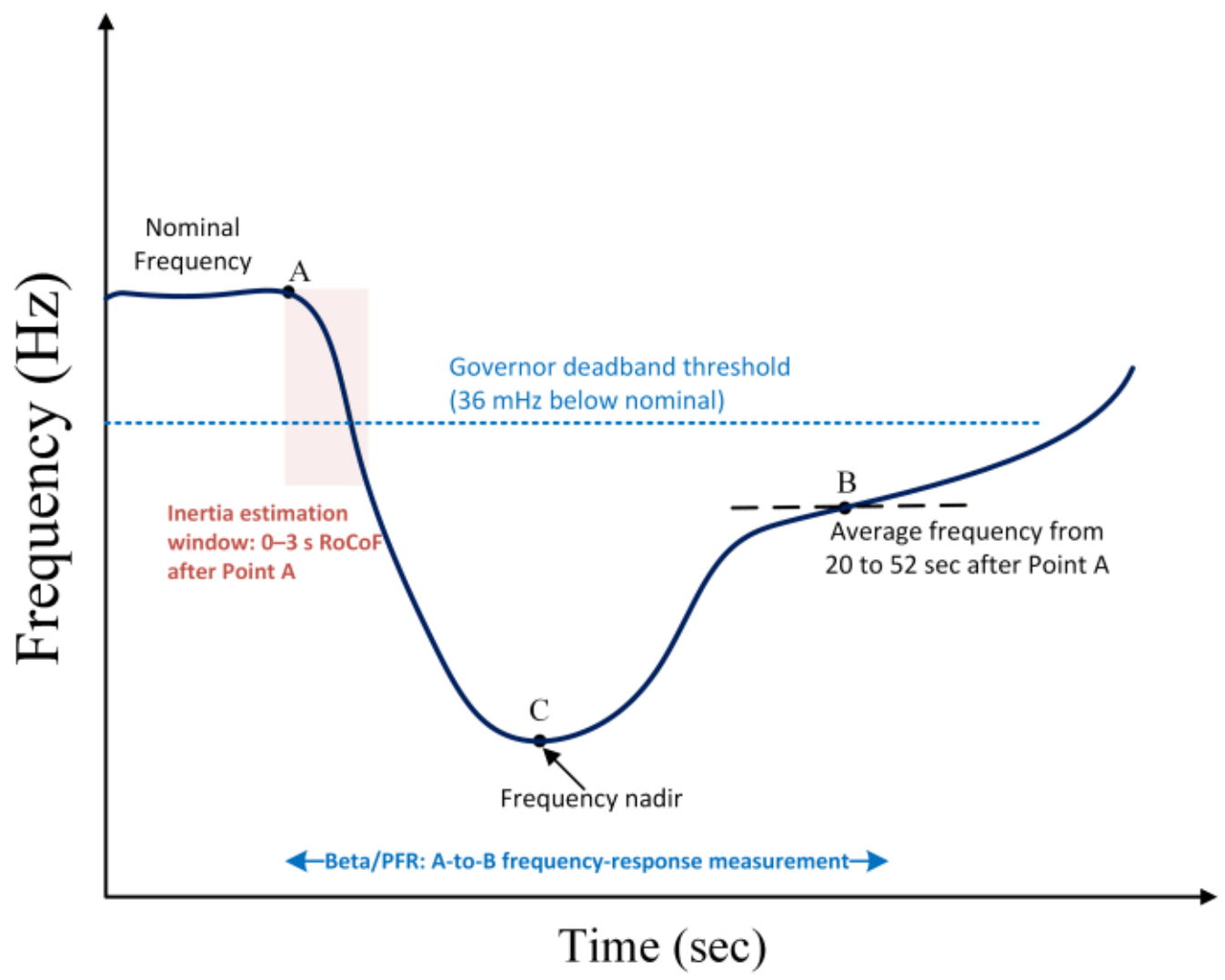

Figure 3: Conceptual frequency response after a generator trip showing the early 0–3 s RoCoF window used for inertia estimation, the governor deadband threshold, and the A-to-B frequency deviation used for beta/PFR estimation, where $f_B$is averaged from 20 s to 52 s after Point A.

### 3.4 Quarterly Inertia Trend Estimation

To characterize the long-term evolution of inertia, the screened disturbance events are grouped by quarter, and the quarterly inertia is estimated from the MW–RoCoF relationship using a least-squares fit. If the vectors of event RoCoF values and disturbance MW values within a quarter are denoted by RoCoF and MW, respectively, the fitted inertia-related coefficient can be obtained as:

$$2\widehat{H} = (RoCoF^T RoCoF)^{-1} \cdot RoCoF^T \cdot MW \tag{5}$$

and the corresponding inertia in MVA·s can be expressed as:

$$\widehat{M} \ = \ (RoCoF^{\mathrm{T}} RoCoF)^{-1}. RoCoF^{\mathrm{T}}. MW \ \cdot \ (f_{syn}/\ 2) \tag{6}$$

where $f_{syn}$ is the synchronous frequency. The quarterly values obtained through this process are then used to construct the long-term EI inertia trend discussed in Section 4. Quarterly estimation is adopted to capture seasonal variation in inertia while maintaining a sufficient number of events in each estimation interval.

### 3.5 PFR Estimation Using the Beta Metric

The primary frequency response characteristic, often denoted by $\beta$ and known as beta, is commonly interpreted as the aggregate frequency response of the system in MW/0.1 Hz [1]. It is evaluated from disturbance size and frequency deviation using event-based measurements as first introduced in [8]. In NERC terminology, beta, represents the total initial response of the system to a frequency excursion, including inertial response, governor-based primary control, and load response. In this paper, $\beta$ is estimated using a least-squares formulation based on the relationship between event frequency deviation and disturbance MW.

$$\Delta f \ = \ f_A \ - \ f_B \tag{7}$$

where $f_A$ is the frequency at the detected disturbance start point and $f_B$ is the average frequency from 20 s to 52 s after Point A. This definition is illustrated in Fig. 3 and is used to improve consistency across events.

For a selected set of events, beta is estimated using least squares as:

$$\hat{\beta} = (\Delta F^{T}\Delta F)^{-1}\Delta F^{T}MW \quad (8)$$

where ΔF is the vector of event frequency deviations. In principle, a larger beta indicates a stronger aggregate primary frequency response, since a larger disturbance is associated with a given frequency deviation. However, the observed beta from field events is not solely a system constant. It is affected by the operating condition at the time of the disturbance, especially the pre-disturbance frequency and the event MW size. Therefore, the beta trend is evaluated using event conditioning to reduce the influence of these event-dependent effects.

### 3.6 Event Conditioning for Beta Trend Analysis

Unlike the RoCoF-based inertia estimate, the β metric is highly sensitive to pre-event operating conditions and therefore exhibits substantial event-to-event scatter. Prior measurement-based analysis [4, 8] for the EI has shown that β has a negative correlation with pre-disturbance frequency and a positive correlation with event MW size. This behavior is consistent with the role of governor deadbands. When the system starts from a higher pre-disturbance frequency, more generators remain within their deadbands during the early stage of frequency decline. This reduces the exhibited primary response and leads to a lower observed β. In contrast, when the pre-disturbance frequency is lower, governors are activated earlier, which tends to strengthen the exhibited response. Likewise, larger events generally reveal stronger β more consistently than smaller disturbances. Fig. 4 illustrates these relationships using EI event data from 2013 to 2025.

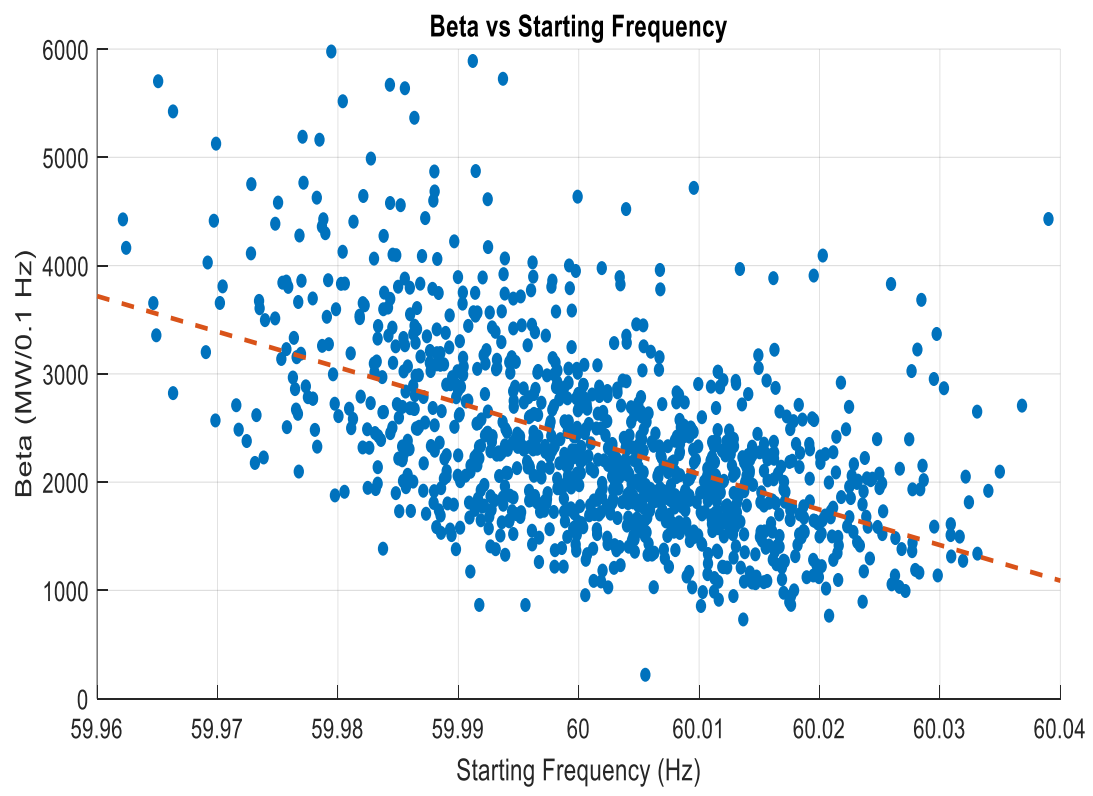

(a)

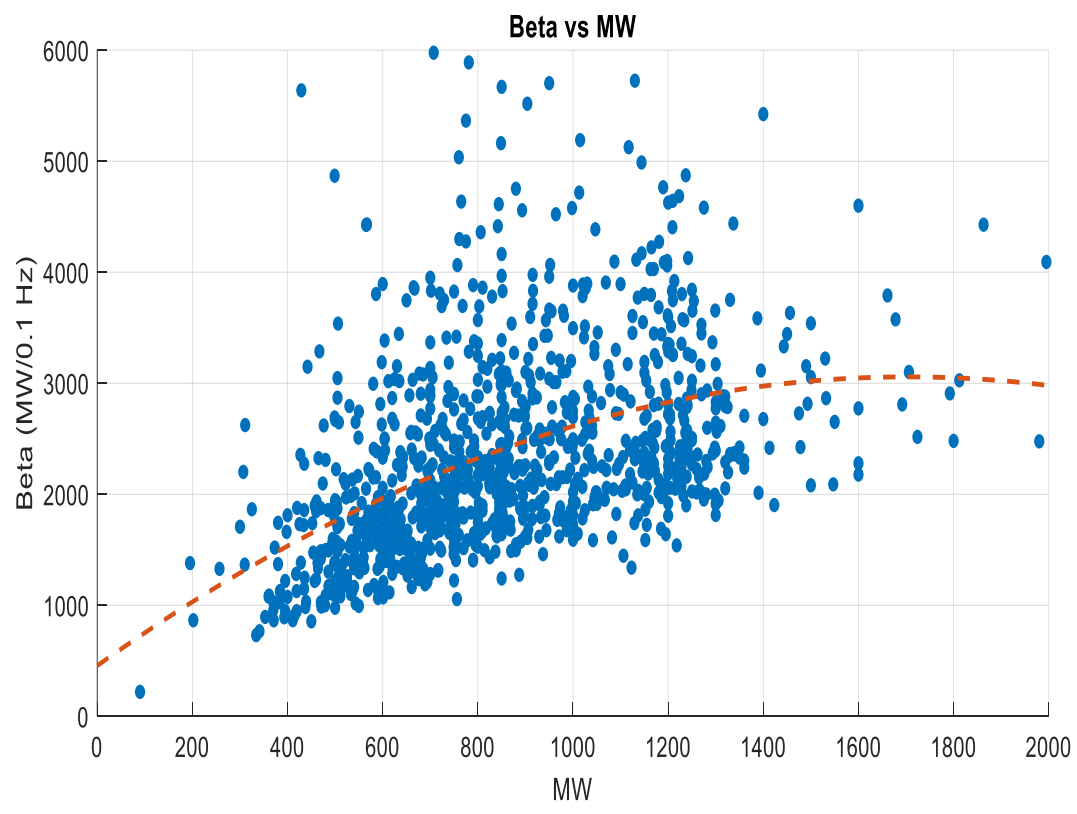

(b)

Figure 4: Dependence of beta on event operating conditions for EI generation-trip events from 2013 to 2025: (a) beta versus starting frequency and (b) beta versus disturbance size in MW.

Accordingly, the β-trend analysis in this paper is developed in two stages. First, all qualified events are used to examine the dependence of β on event conditions, particularly starting frequency and disturbance MW. Second, the event-filtering criteria proposed in prior PFR work [4] are applied to select events within comparable starting-frequency and disturbance-size ranges. In this study, events satisfying starting frequency below 59.99 and event size greater than 800 MW are retained for the annual β trend calculation. This filtering improves the correlation between MW and $\Delta f$. As a result, this reduces the influence of event-dependent operating conditions and enables a fairer year-to-year comparison of the exhibited aggregate frequency response.

The methodology described above yields one quarterly inertia series and one filtered annual β series, which are presented and interpreted jointly in Section 4.

## 4. RESULTS AND DISCUSSION

This section presents the inertia and PFR results obtained for EI and discusses their joint implications for long-term frequency performance. The results are organized in four parts. First, the quarterly inertia trend is presented together with the fuel-mix evolution used to support its interpretation. Second, the annual beta trend obtained using all qualified events is presented as a baseline reference. Third, the filtered annual beta trend is presented as the more comparable long-term PFR indicator. Finally, the inertia and beta trends are interpreted jointly to assess how frequency performance in the EI has evolved over the study period.

### 4.1 Inertia Trend in the Eastern Interconnection

Fig. 5(a) shows the quarterly inertia trend estimated from the screened EI disturbance events using the selected 3-second RoCoF window. Several features can be observed. First, the inertia exhibits noticeable seasonal variation, with higher values generally appearing during periods of stronger system loading and generator commitment. This seasonal behavior is consistent with the expectation that a larger number of online synchronous units contribute more effective inertial support to the interconnection. Second, the long-term trend is not characterized by a uniform monotonic decline. Instead, the trend shows intervals of reduction, relatively steady behavior, and modest recovery, indicating that the evolution of EI inertia has been shaped by multiple offsetting factors rather than by renewable penetration alone.

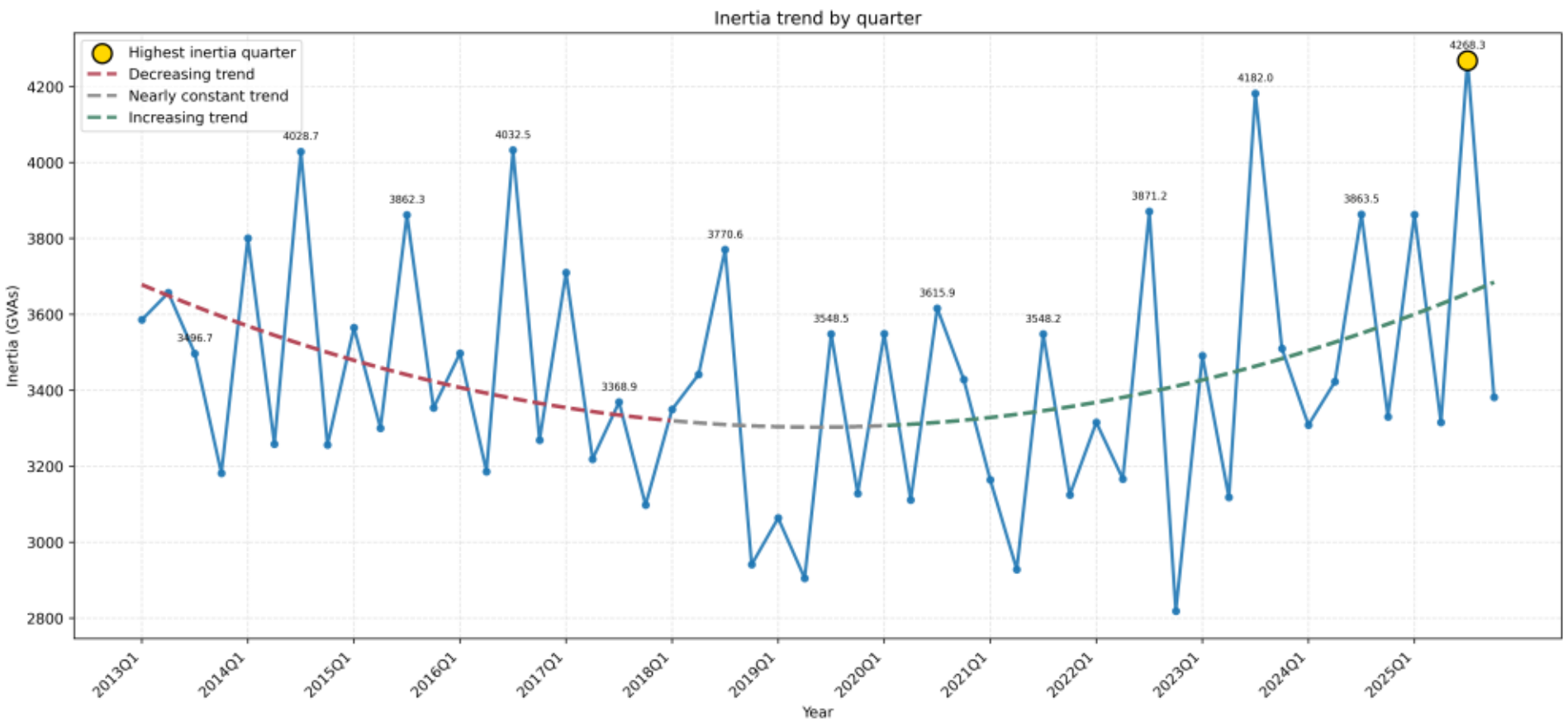

(a)

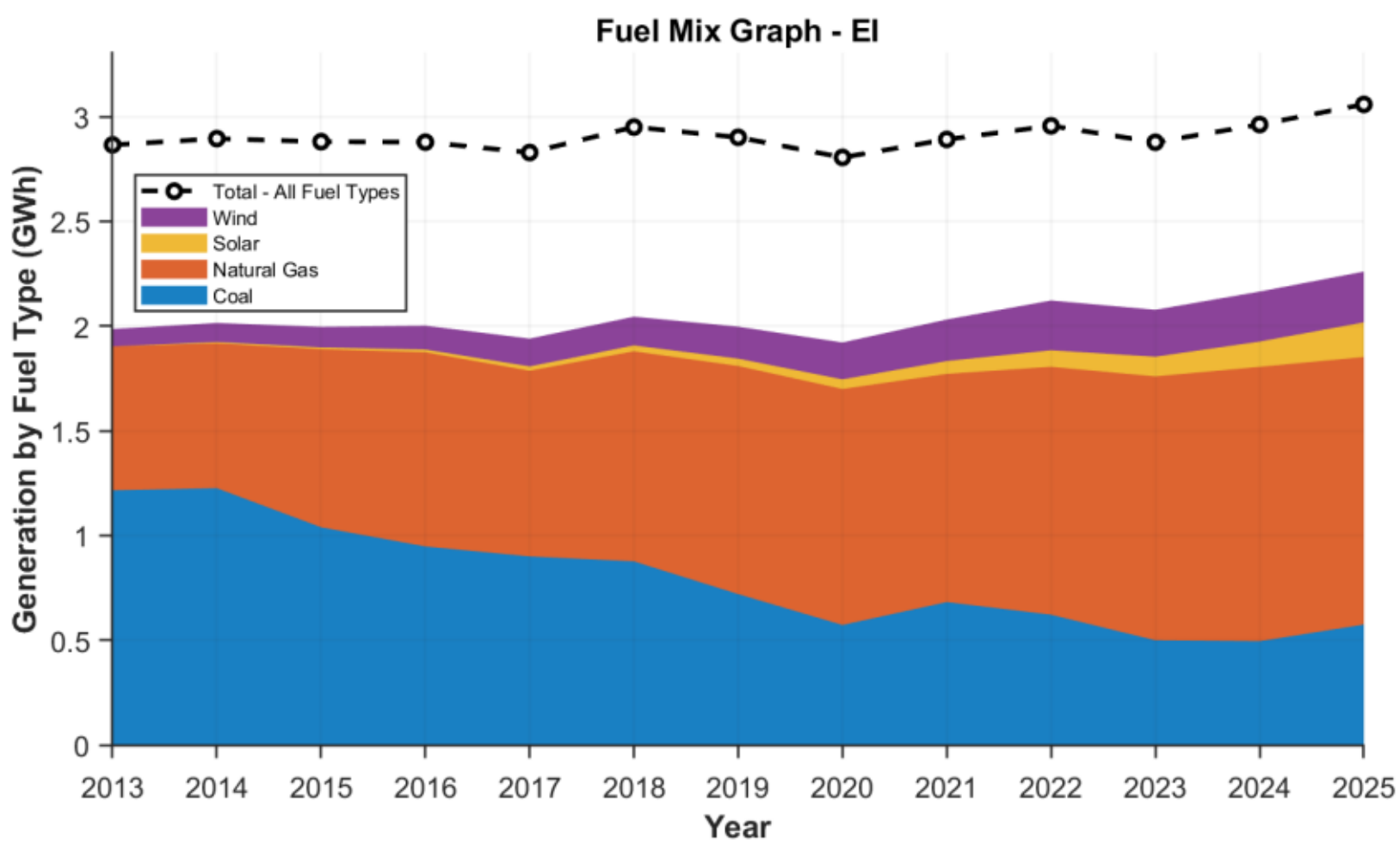

(b)

Figure 5: Long-term inertia and generation-mix trends in the Eastern Interconnection: (a) quarterly inertia trend estimated from screened EI disturbance events using the selected 3 s RoCoF window, and (b) annual EI generation by fuel type based on U.S. Energy Information Administration data [9].

To support interpretation of this trend, Fig. 5(b) presents the corresponding long-term evolution of EI generation by fuel type, obtained from the U.S. Energy Information Administration (EIA) data [9]. Over the study period, coal generation decreases substantially, while natural-gas generation increases and wind and solar continue to expand. A simple expectation might be that increasing inverter-based penetration would lead directly to a strong decline in effective inertia. However, the measured EI inertia trend suggests that this interpretation is incomplete. The increase in gas-fired generation, together with seasonal commitment effects and continued operation of synchronous resources, appears to have partially offset the inertia reduction that might otherwise be expected from increased renewable integration alone. Thus, the measured inertia trend should be interpreted as the net result of changing resource mix, commitment patterns, and operating conditions.

It is also important to recall that the inertia estimate used herein represents an effective disturbance-based interconnection inertia. Therefore, it reflects the actual system behavior observed during events rather than only the physical rotating kinetic energy of online synchronous machines.

### 4.2 Filtered Beta Trend

Using the event-conditioning approach described in Section 3.6, the annual beta trend is estimated after applying the filtering criteria. Specifically, events satisfying $f_A < 59.99$ and MW > 800 are retained for the trend calculation. The starting-frequency filter selects events for which governor response is expected to be more consistently exhibited, while the disturbance-size filter removes smaller events that are more sensitive to measurement scatter and deadband-related effects. The resulting filtered annual beta trend is shown in Fig. 6.

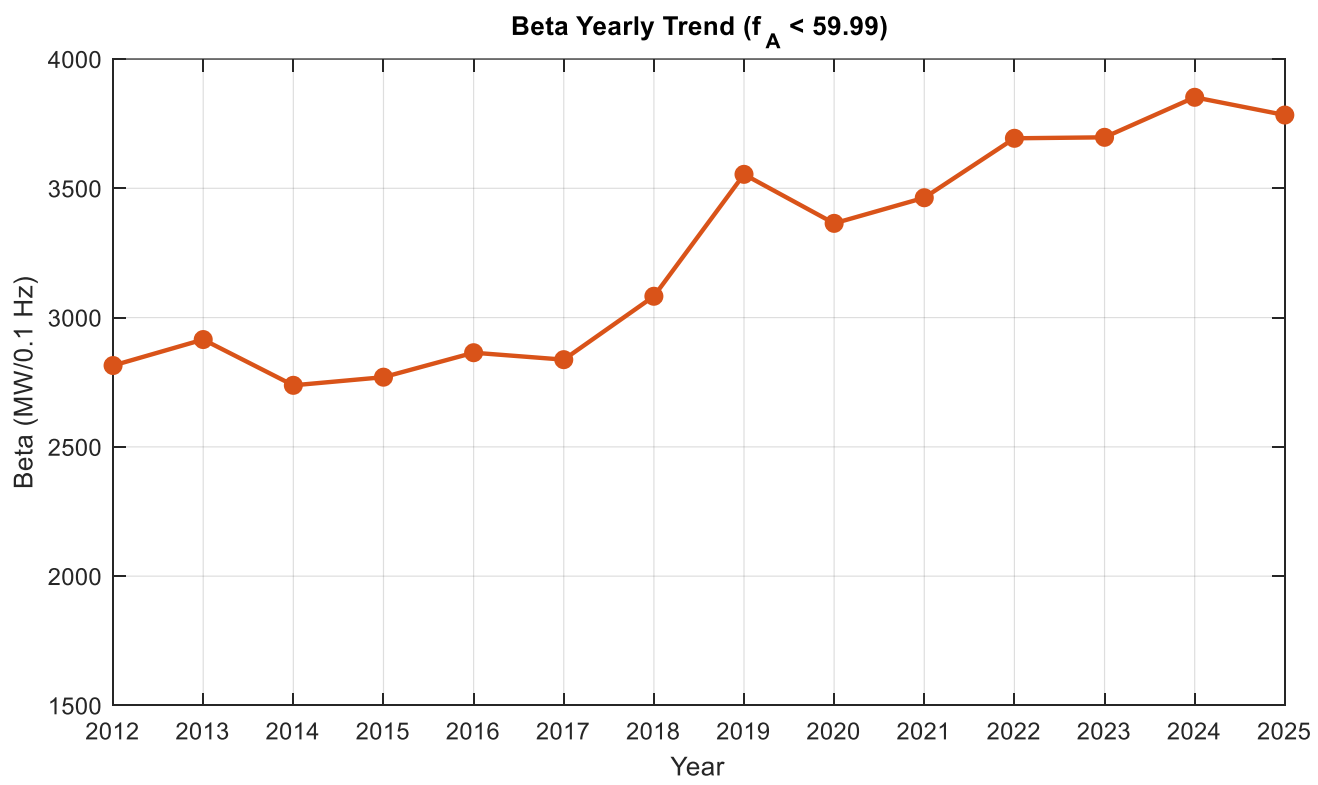

Figure 6: Filtered Beta Trend

One possible explanation for the increasing beta trend is the changing composition of governor-responsive generation. As shown in Fig. 5(b), natural-gas generation increased while coal generation declined over the study period. This increase occurred during a period of substantial combined-cycle expansion. EIA reported that almost 90% of the natural-gas-fired capacity added in the United States in 2018 consisted of combined-cycle generators [10]. NERC further notes that many combustion turbines in combined-cycle configurations have a 4% droop setting, compared with a recommended maximum of 5% for the steam-turbine portion of combined-cycle plants and other generating units [11]. Since governor response is inversely related to droop, increased participation of lower-droop combustion-turbine resources within the growing combined-cycle fleet may have contributed to the higher beta values observed in later years. This interpretation should be treated as a system-level explanation rather than direct unit-level causal proof.

### 4.3 Joint Interpretation of Results

The inertia and beta trends provide different views of how EI frequency performance has changed over the study period. When the two results are considered together, they show that the EI disturbance response has not followed a simple one-directional pattern. The effective inertia trend is non-monotonic, while the filtered beta trend shows a generally stronger aggregate response in the later part of the study period. This indicates that changes in early inertial behavior do not directly translate into proportional changes in the broader frequency response.

This result is important because both metrics are derived from disturbance events, but they emphasize different parts of the response. Inertia represents the early RoCoF-based behavior immediately after the disturbance, while beta reflects the aggregate response over the selected post-disturbance interval. Therefore, beta can change because of factors beyond inertia, including governor response, governor deadbands, load-frequency sensitivity, event size, starting frequency, and operating conditions. This helps explain why the beta trend can improve or remain strong even when the inertia trend is variable.
The joint result suggests that EI frequency performance has been supported by multiple interacting mechanisms. Resource-mix transition provides the background context, but the measured trends indicate that system response is also shaped by synchronous resource commitment, gas-fired generation, governor-responsive resources, load behavior, and event operating conditions. These factors can offset, reinforce, or reshape the effects that would be expected from resource-mix change alone.

The key insight is that the EI disturbance response appears to have been broadly maintained during the study period, but not through a fixed or unchanged response mechanism. Effective inertia did not decline uniformly, and the filtered beta trend indicates stronger aggregate frequency response in several later years. Therefore, the main conclusion is not that EI frequency performance has simply improved or degraded, but that different components of frequency performance have evolved differently. This supports the need to interpret inertia and beta jointly when assessing long-term frequency behavior in the EI.

## 5. CONCLUSION

This paper presented a measurement-based assessment of long-term inertia and beta trends in the U.S. Eastern Interconnection using FNET/GridEye frequency measurements and confirmed disturbance-event data. The results show that EI frequency performance has not evolved in a simple one-directional manner. The effective inertia trend is seasonal and non-monotonic, while the filtered beta trend suggests a generally stronger aggregate response in several later years. These findings indicate that the EI disturbance response has been broadly sustained during the generation-mix transition, but through changing combinations of synchronous resource commitment, gas-fired generation, governor-responsive behavior, load response, and operating conditions. Overall, the study shows that joint interpretation of inertia and beta provides a more complete understanding of interconnection-level frequency performance than either metric alone. One limitation of the beta analysis is that the number of NERC-confirmed disturbance events is limited, and the event-conditioning filters further reduce the number of events available for annual estimation. With a larger set of qualified disturbance events, the filtered beta trend could be estimated with greater statistical confidence and reduced year-to-year uncertainty. Future work will therefore focus on expanding the qualified event database and evaluating the sensitivity of beta trends to the adopted filtering criteria.

**ACKNOWLEDGEMENT**
This material is based upon work supported by the US Department of Energy, Office of Electricity (OE) under contract DE-AC05-00OR22725.